# Magnetic Fields of Extrasolar Planets: Planetary Interiors and Habitability

*A white paper submitted to the National Academy of Science Committee on Exoplanet Science Strategy*

J. Lazio[1] (Jet Propulsion Laboratory, California Institute of Technology)

G. Hallinan (California Institute of Technology),

V. Airapetian (NASA/GSFC), D. A. Brain (University of Colorado, Boulder), C. F. Dong (Princeton University), P. E. Driscoll (Carnegie Institute for Science), J.-M. Griessmeier (LPC2E-Universitè d'Orlèans/CNRS, Station de Radioastronomie de Nançay, Observatoire de Paris), W. M. Farrell (NASA/GSFC), J. C. Kasper (University of Michigan), T. Murphy (University of Sydney), L. A. Rogers (University of Chicago), A. Wolszczan (Pennsylvania State University), P. Zarka (Observatoire de Paris, CNRS, PSL), M. Knapp (MIT EAPS), C. R. Lynch (University of Sydney), J.D. Turner (University of Virginia)

[1] 4800 Oak Grove Dr, M/S 67-201, Pasadena, CA 91109; 818-354-4198; Joseph.Lazio@jpl.nasa.gov

Jupiter's radio emission has been linked to its planetary-scale magnetic field, and spacecraft investigations have revealed that most planets, and some moons, have or had a global magnetic field. Generated by internal dynamos, magnetic fields are one of the few remote sensing means of constraining the properties of planetary interiors. For the Earth, its magnetic field has been speculated to be partially responsible for its habitability, and knowledge of an extrasolar planet's magnetic field may be necessary to assess its habitability. The radio emission from Jupiter and other solar system planets is produced by an electron cyclotron maser, and detections of extrasolar planetary electron cyclotron masers will enable measurements of extrasolar planetary magnetic fields.

This white paper draws heavily on the W. M. Keck Institute for Space Studies report *Planetary Magnetic Fields: Planetary Interiors and Habitability* (Lazio, Shkolnik, Hallinan, et al. 2016), it incorporates topics discussed at the American Astronomical Society Topical Conference "Radio Exploration of Planetary Habitability," it complements the Astrobiology Science Strategy white paper "Life Beyond the Solar System: Space Weather and Its Impact on Habitable Worlds" (Airapetian et al. 2018), and it addresses aspects of planetary magnetic fields discussed in the 2015 NASA Astrobiology Strategy.

## Scientific Background

Even early explanations for Earth's magnetic field linked it to Earth's interior structure. After the discovery of Jupiter's radio emission (Burke & Franklin 1955; Franklin & Burke 1956), it was determined that this radiation was linked to Jupiter's magnetic field (Carr & Gulkis 1969), which was then tied to the planet's interior structure. Today, remote sensing and *in situ* measurements have shown that the Earth, Mercury, Ganymede, and the giant planets of the solar system all contain internal dynamos that generate planetary-scale fields; Mars and the Moon show residual magnetism indicative of past dynamos.

Internal dynamos arise from differential rotation, convection, compositional dynamics, or a combination of these processes. Knowledge of extrasolar planetary magnetic fields places constraints on internal compositions and dynamics, which will be difficult to determine by other means, as well as informs the extent to which the surfaces and atmospheres of terrestrial planets are shielded and potentially habitable.

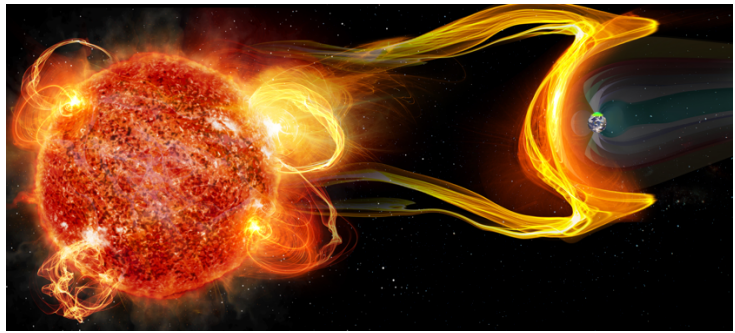

Figure 1. Artist's impression of an extrasolar planet's magnetosphere shielding the planet from a violent eruption of its host low-mass star. The Earth, Mercury, Ganymede, and the giant planets in the solar system generate magnetospheres from internal currents, which provide information about their interior structures. Magnetosphere-stellar wind interactions produce radio emissions that may be detectable over interstellar distances. (Figure not to scale; credit: Keck Institute for Space Studies)

The stellar wind, a supersonic magnetized plasma, incident on a planet's magnetosphere (Figure 1), is an energy source to the magnetosphere. **An electron cyclotron maser, resulting from a magnetosphere-solar wind interaction, has been detected from the Earth and all of the gas giants in the solar system.**

## Planetary Interiors

The detection of even a single extrasolar planetary magnetic field could provide essential information to extend our knowledge of planetary interiors and dynamos. A limiting factor in understanding planetary dynamos is the small sample in the solar system (Stevenson 2010; Schubert & Soderlund 2011). Just as the discoveries of hot Jupiters gave crucial insights to the diversity of planets, the detection of extrasolar magnetic fields will improve our knowledge of planetary interiors and magnetic dynamos, including in our solar system.

Figure 2 illustrates that inferring planet compositions is an under-constrained inversion problem, because planets with disparate compositions can have identical masses and radii. GJ 1214b (6.5 $M_\oplus$, 2.7 $R_\oplus$) provides an example—it could have a rock-ice interior and primordial H/He envelope or be a water planet (> 47% $H_2O$ by mass) shrouded in vapor from sublimated ices or a super-Earth harboring a H-rich outgassed atmosphere.

Measurements of extrasolar magnetic fields will add an extra dimension with which to characterize planets. Magnetic field measurements, providing information about interior structures and compositions, will complement measurements of upper atmosphere compositions obtained by transit transmission spectroscopy from ground-based telescopes and missions such as TESS. The planet mass-radius diagram will extend to a mass-radius-field strength diagram because a magnetic field requires an internal, convecting, electrically conducting fluid.

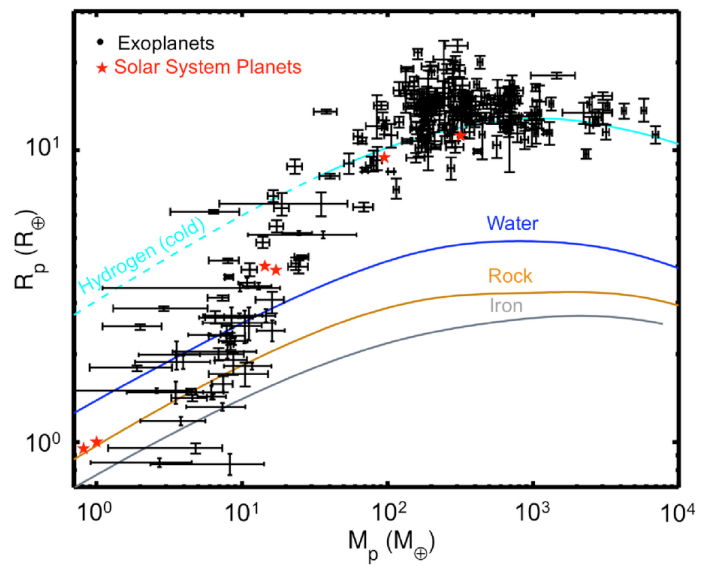

Figure 2. Mass-radius relations of extrasolar planets have degeneracies, with multiple potential compositions describing a planet. Curves show the mass-radius relation for a planet with a pure composition (hydrogen, water, rock, or iron). Mixes of different constituents also satisfy the mass-radius relation, but magnetic field measurements provide constraints that can break these degeneracies. (Rogers 2016)

This presence of convection might be most informative for rocky planets, which can, but are not guaranteed to, have electrically conducting liquid iron cores. If the oxygen fugacity is too high during its assembly, a planet's iron may be oxidized and incorporated into the silicate mantle, instead of being differentiated into a core (Elkins-Tanton & Seager 2008). Partial solidification of the core may limit the range of planet masses with sufficiently liquid cores to sustain dynamos, and the extent to which an iron core solidifies is also sensitive to the presence of volatiles. Further, the energy budget for convection in Earth's core is marginal. Higher equilibrium temperature (> 1500 K), stronger tidal heating, higher concentrations of radio nuclei, the presence of a thick H/He envelope, or a stagnant lid tectonic regime could turn off convection (and a dynamo) in the core of an otherwise Earth-like planet. The inference of convection via a magnetic field measurement would constrain the planet's thermal evolution and energy budget and may serve as an indirect indication of plate tectonics.

For ice giants, water is electrically conducting above a few thousand Kelvin (e.g., French et al. 2009). A straightforward prediction is that Neptune-like planets should sustain planetary-scale dynamos and that the detection of their magnetic fields would confirm their compositions as being substantially water.

Finally, in Jovian planets with massive H/He envelopes, hydrogen is metallic above about 25 GPa (Wigner & Huntington 1935) and they are expected to be convective at depth. The *absence* of magnetic fields would challenge our understanding of the interior structures of giant planets.

## Planetary Habitability

Galactic cosmic rays, solar (stellar) energetic particles (SEPs), and UV radiation produce harsh space environments around the terrestrial planets; similar conditions are expected for extrasolar planets around middle-aged main sequence F, G, and K stars, even more harsh conditions are likely around young stars or M dwarfs. While SEPs and UV radiation may play a role in the origin of life (e.g., Lingam et al. 2018), in the long term, if directly incident on a surface, the radiation and high-energy particles are expected to be harmful (destructive) to Earth-like biological tissue; even if not incident on the surface, their effects may contribute to erosion of a planet's atmosphere.

A comparison of Earth to Mars and Venus is often used to support the argument that magnetic fields can prevent loss of water from atmospheres (e.g., Lundin et al. 2007). Earth, with its strong dipole field, has an atmosphere that allows liquid surface water and sustains life. Mars, which lacks a strong global field, has an atmospheric pressure less than 1% that of Earth, but surface magnetization and morphology such as lake beds suggest that a strong global magnetic field and surface liquid water existed about 4 Gyr ago. The Venusian atmosphere, unprotected by a global magnetic field, has a surface pressure 90× Earth, but with little water content. Early water on Venus could have been disassociated, with the hydrogen lost to space and the oxygen absorbed into crustal rocks (e.g., Driscoll & Bercovici 2013). Indeed, Earth would have a substantial $CO_2$-dominated atmosphere were it not for the effect of Earth's oceans in removing $CO_2$ from the atmosphere.

Spacecraft observations confirm that the solar wind stagnates at the bow of a planet's magneto-sphere, with the bulk of the plasma deflected around the magnetospheric cavity. As such, it seems plausible that a global field reduces a planet's atmospheric loss, in particular helping to retain the hydrogen and oxygen ions that make up water, yet, surprisingly, Venus, Earth, and Mars have similar present-day atmospheric losses ($\sim 10^{25}$ $s^{-1}$). Studies of the terrestrial polar regions, where the solar wind is directly accessible via connecting field lines, indicates substantial atmospheric ($O^+$) loss (Moore & Khazanov 2010). **Ideally, a large sample of planets, with a range of** **atmospheric compositions and magnetic field properties would be available to test the extent to which the presence of a magnetic field protects an atmosphere.**

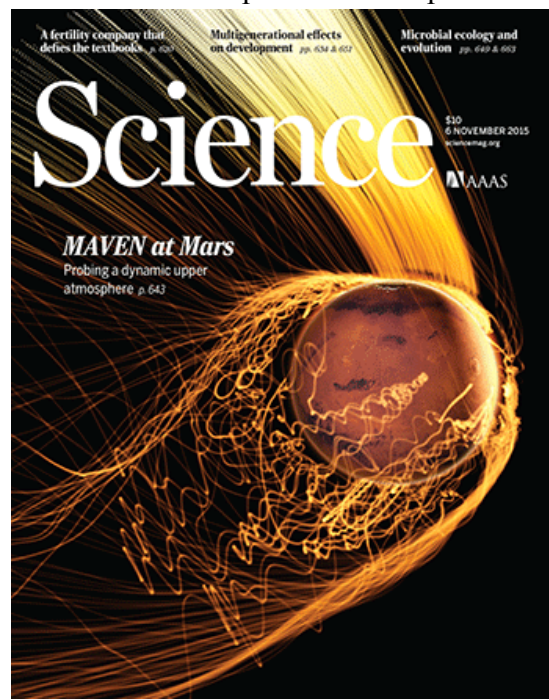

Figure 3. The Mars Atmosphere and Volatile Evolution (MAVEN) mission has provided dramatic evidence for atmospheric erosion. Unshielded from the solar wind due to the lack of a global magnetic field, Mars' atmosphere is eroded via an interaction with the solar wind. Assessing extrasolar planetary magnetic fields may be essential in understanding their potential habitability. (From Jakosky et al. 2015; reprinted with permission from AAAS.)

## Progress in the Next Decade

A direct measure of a planet's magnetic field can be determined from its electron cyclotron maser emission. This emission occurs up to a characteristic frequency determined by the polar cyclotron frequency, which in turn depends upon the planet's magnetic field. Scaling laws based on the solar system planets exist (Farrell, Desch, & Zarka 1999; Zarka et al. 2001; Christensen 2010), with $f_{ECM} = 2.8$ MHz $B$, for a field in Gauss; for Jupiter, $f_{ECM,J} \approx 30$ MHz. These scaling relations are *predictive*, with the luminosities of Uranus and Neptune predicted *before* the Voyager 2 encounters (Desch & Kaiser 1984; Desch 1988; Millon & Goertz 1988).

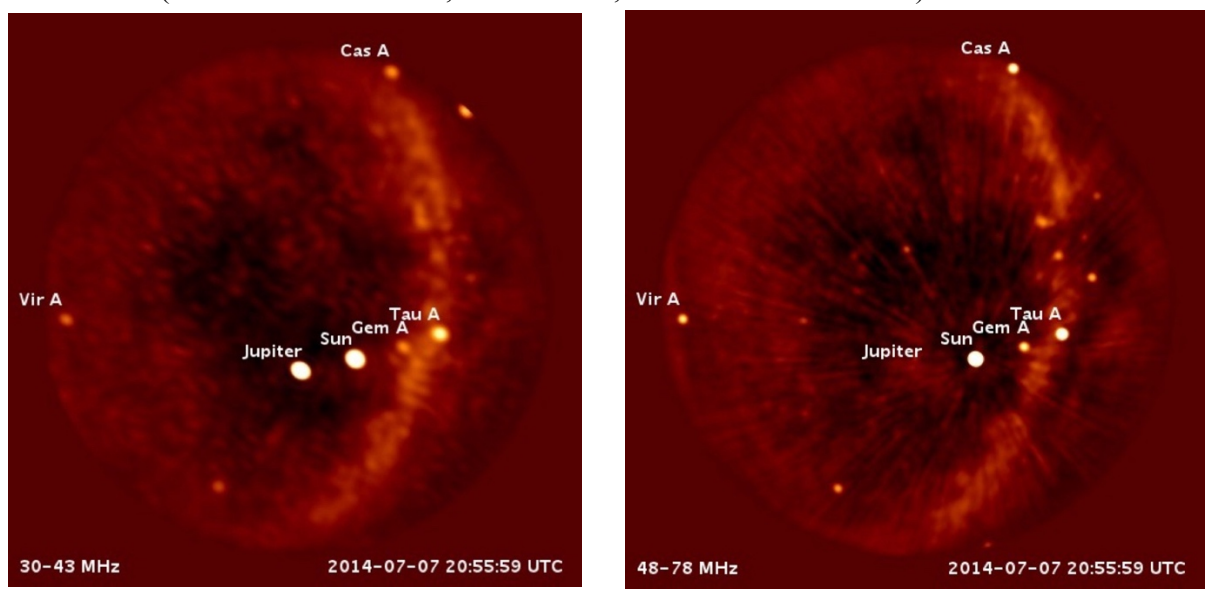

Figure 4. Jupiter as an extrasolar planet, as observed by the Long Wavelength Array-Owens Valley Radio Observatory in the bands 30 MHz–43 MHz (*left*) and 47 MHz–78 MHz (*right*). Strong sources are labeled, notably including Jupiter and the Sun. The absence of Jupiter in the higher frequency image is consistent with the cutoff of electron cyclotron maser emission where the local plasma frequency exceeds the local cyclotron frequency within the planet's magnetosphere. Ground-based telescopes have been making steady progress toward detecting analogous emissions from nearby giant planets; a space-based telescope would be required to study planets with weaker fields, such as ice giants or terrestrial planets and might be expected for mini-Neptunes and super-Earths in other solar systems. (Credit: M. Anderson)

Based on the scaling laws, detecting planetary magnetospheric emissions is most promising at frequencies below 90 MHz, though young, hot extrasolar planets may have fields strong enough to be detectable at higher frequencies. Murphy et al. (2015) and Lazio et al. (2016) provide summaries of the current observations. No extrasolar planetary electron cyclotron maser has been unambiguously detected, but a combination of limited sensitivity and frequency coverage is probably responsible—the most sensitive searches below 90 MHz have been at 74 MHz, which can be compared with Jupiter's cutoff frequency near 30 MHz.

The past decade has witnessed the initial operation of telescopes designed to observe below 90 MHz, including multiple instances of the Long Wavelength Array (LWA, New Mexico and California) and the Low Frequency Array (LOFAR, the Netherlands). A significant constraint to ground-based telescopes is the Earth's ionosphere, which is opaque below about 10 MHz. This natural limit prevents ground-based observations of the Earth itself(!), Saturn, Uranus, and Neptune; plausibly, this limit precludes observations of super-Earths and mini-Neptunes. The Sun Radio Interferometer Space Experiment (SunRISE) concept, in NASA Heliophysics Phase A, is a space-based telescope designed to observe the Sun at frequencies below 15 MHz. SunRISE is unlikely to be sensitive enough to detect an extrasolar planet, but it should detect Saturn, thereby proving the concept of a space-based telescope to study extrasolar magnetospheric emissions.

A natural strategy thus presents itself for the next decade and beyond.

- Ground- and space-based studies of the solar neighborhood will refine the target list of extra-solar planets for which magnetic field measurements would be both possible and valuable, particularly as they might relate to atmospheric composition and structure.
- Ground-based telescopes, e.g., the LWA and LOFAR, will improve upon the sensitivity and techniques for detecting extrasolar planetary magnetospheric emissions, with a likely focus on giant planets, and potential surprises from mini-Neptunes, if their fields are sufficiently strong.
- The Juno mission and subsequent outer planet missions will improve our knowledge of the magnetic dynamos of gas giant planets.
- SunRISE would prove the technologies for a future space-based telescope designed to study extrasolar planets with lower magnetic field strengths, such as mini-Neptunes, super-Earths, and terrestrial planets.